# Distinguishing Hot-Electron and Optomechanical Pathways at Metal-Molecule Interfaces

Bing Gao[1], Jameel Damoah[1], Wassie M. Takele[1,†], and Terefe G. Habteyes[1]*

[1]Department of Chemistry & Chemical Biology, and Center for High Technology Materials, University of New Mexico; Albuquerque, New Mexico, USA.

*Corresponding author. Email: habteyes@unm.edu.

[†]Present address: Department of Physics, University of Georgia; Athens, USA.

**Abstract:** Energy and charge transfer between molecules and metal surfaces underpin heterogeneous catalysis, surface-enhanced spectroscopies, and plasmon-driven chemistry, yet the microscopic origins of vibrational excitation at metal interfaces remain unresolved. Here we use temperature-dependent surface-enhanced Raman scattering (SERS) to directly distinguish plasmon–vibration optomechanical coupling from hot-electron–driven excitation. By probing thionine adsorbed on gold nanostructures at 295 K and 3.5 K, we show that pronounced anti-Stokes scattering at cryogenic temperature arises from optical pumping of vibrational populations, whereas room-temperature spectra are governed by thermal population. Bromide co-adsorbates play a decisive role by guiding molecular alignment, inducing surface atom displacements, and enabling transient adsorption geometries that activate otherwise Raman-inactive vibrational modes. In the absence of bromide, distinct excitation pathways emerge, reflecting competition between optomechanical coupling and charge-transfer processes associated with molecular polarization along the optical field or orientation relative to the metal surface. These results establish molecular optomechanics as a sensitive probe of surface–molecule interactions and demonstrate how anion-mediated surface dynamics regulate energy flow at plasmonic interfaces.

Understanding how molecules interact with metal surfaces is a central problem in surface science, as these interactions determine charge and energy transfer pathways, and potential energy landscapes that ultimately control heterogeneous catalysis, surface-enhanced spectroscopies, chemical sensing, and other applications (*1-3*). A variety of spectroscopic techniques, including infrared (*4*), ultraviolet (*5*), visible, and X-ray spectroscopies (*6*), as well as sum-frequency generation spectroscopy (*7*), can be used to determine elemental composition, identify binding energies and vibrational modes of surface-bound molecules, and assess molecular orientation (*8*). However, these methods often fall short in capturing nuanced variations in adsorption and orientation dynamics due to limited sensitivity at the sub-monolayer and single-molecule levels (*9-12*).

Surface-enhanced Raman scattering (SERS) spectroscopy offers single-molecule sensitivity (*13-15*), enabling direct interrogation of how co-adsorbates and defects mediate surface–molecule interactions and catalytic behavior. In particular, the relative intensity of anti-Stokes SERS provides a direct measure of vibrational population to elucidate the mechanisms of vibrational excitation on metal surfaces, which is central to understand heterogeneous catalysis. Recent progress has led to two distinct mechanistic pictures: vibrational excitation driven by quantum mechanical optomechanical plasmon–vibration coupling (*16, 17*), and excitation mediated by hot electrons generated during plasmon decay (*18-21*). Molecular optomechanics describes a dynamical quantum interaction in which specific vibrational modes can be selectively excited by tuning the excitation frequency relative to the plasmon resonance or vice versa. By contrast, hot-electron–driven excitation is an incoherent process involving transient charge transfer to the adsorbate, followed by return of the electron to the metal leaving the molecule in an excited vibrational state. Although both mechanisms are plausible, identifying the conditions under what condition each dominates would represent a major advance in our mechanistic understanding of surface–molecule interactions.

This work not only identifies the two mechanisms but also leverages molecular optomechanics to probe anion-guided molecular alignment and surface atom dynamics. Bromide anions, which inherently adsorb on plasmonic gold nanorods during synthesis in the cetyltrimethylammonium bromide (CTAB) stabilizing surfactant, are used here as a model co-adsorbate to mediate surface–molecule interactions. Vibrational pumping is studied using SERS measurement at two limiting temperatures, 295 K and 3.5 K, using thionine as a Raman probe. Thionine is chosen as a probe molecule because it is chemically stable and interacts favorably with metal surfaces (*22*). Unlike other phenothiazinium dyes such as methylene blue that undergoes photochemical *N*-demethylation, thionine remains intact under laser illumination (*23-25*). Furthermore, while thiol-containing molecules form direct covalent type bonds with noble metals, thionine adsorbs mainly through dispersion forces (*22*). This noncovalent binding provides a convenient platform to study how co-adsorbates influence vibrational pumping through optomechanical and/or charge-transfer pathways.

Depending on the surface condition and orientation of the molecule, the Raman polarization (**p**) relative to the near-field vector (**E**) in the plasmonic nanocavity (Fig. 1A) can range from $\theta \approx 0$ (strong dipole-field coupling) to $\theta \approx 90$ (weak coupling). At the excitation wavelength of 633 nm, the lowest-energy electronic transition of thionine adsorbed on metal surfaces is accessible (*26*) (Fig. 1B, fig. S1 and S2), allowing further enhancement of Raman scattering cross-section due to electronic resonance effect (*27*). The Raman-active vibrational mode centered at 479 cm$^{-1}$ is used primarily to study vibrational pumping because of its high signal-to-noise ratio; however, the results are reproducible for other vibrational modes. For this vibrational mode, the anti-Stokes ($I_{aS}$) to Stokes ($I_S$) scattering intensity ratio ($\rho = I_{aS}/I_S$) is essentially zero at temperatures below 50 K (Fig. 1B), indicating a negligible thermal population ($n_{th} \approx 10^{-85}$ at 3.5 K compared to $n_{th} \approx 0.1$ at 295 K).

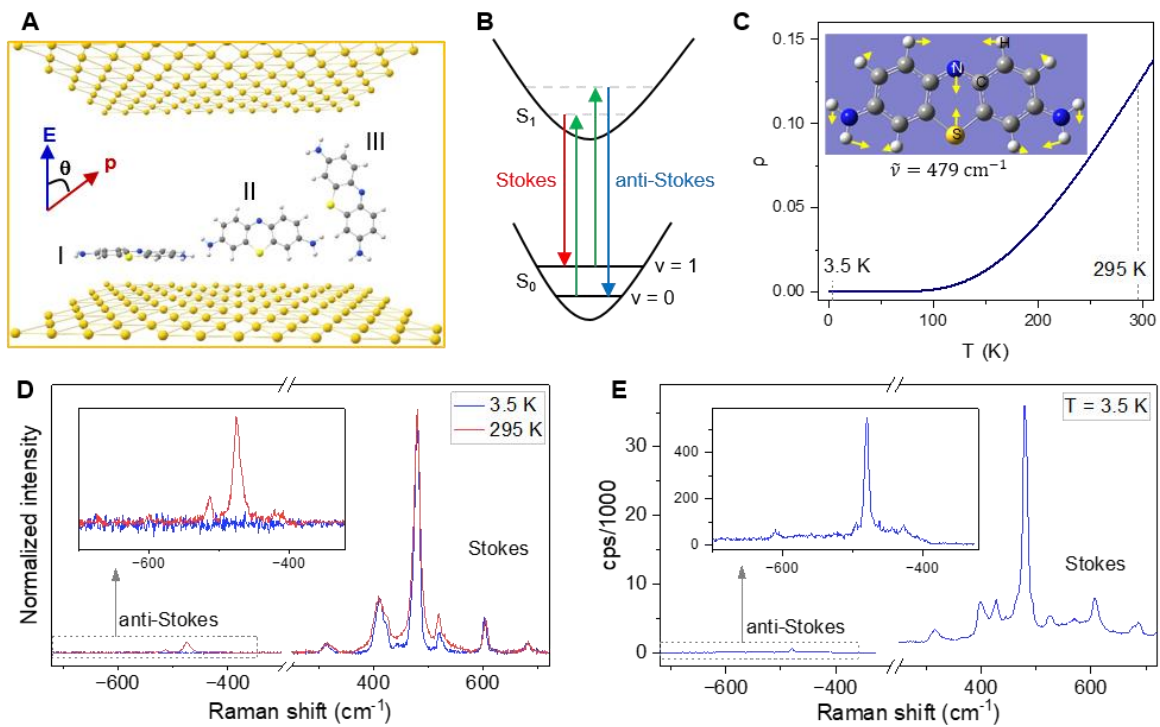

**Fig. 1. Experimental approach and principles**. (**A**) Molecular (thionine) polarization (**p**) in a plasmonic nanocavity with respect the gold surface and the electric field **E**. (**B**) Schematic showing Stokes (S) and anti-Stokes (aS) transitions between singlet ground ($S_0$) and single excited ($S_1$) electronic states. (**C**) Anti-Stokes–to–Stokes intensity ratio based on the thermal vibrational population of the 479 cm$^{-1}$ vibrational mode of thionine. Inset shows the atom displacement vectors for the 479 cm$^{-1}$ vibrational mode. (**D**) Raman spectra of thionine film on silicon wafer at 295 K (red line) and 3.5 K (blue line). The inset shows no vibrational peaks in the anti-Stokes branch at 3.5 K. (**E**) SERS spectrum of thionine on AuNRs at 3.5 K. The plot in the inset shows the observation of anti-Stokes vibrational peaks at 3.5 K due to plasmon pumping of the excited vibrational state.

For a thionine film on silicon in the absence of plasmonic nanoparticles, anti-Stokes Raman peaks are readily observed at 295 K (red trace, Fig. 1D) but vanish at 3.5 K (blue trace, Fig. 1D), as expected from thermal vibrational populations. In sharp contrast, SERS spectra of thionine obtained from aggregates of gold nanorods exhibit pronounced anti-Stokes peaks even at 3.5 K (Fig. 1E), demonstrating the generation of vibrational quanta through surface–molecule interactions rather than thermal excitation. As discussed below, the laser-power dependence of the anti-Stokes Raman signal at 3.5 K is fully consistent with vibrational plasmon pumping and molecular optomechanics theory. A detailed analysis in the Supplementary Materials shows that the observed anti-Stokes signal at 3.5 K cannot be attributed to a temperature increase within the SERS hotspot due to plasmonic heating, particularly given the low laser intensities used in these experiments (36 – 367 μm/μm$^2$). We therefore focus on the mechanism of vibrational excitation by systematically comparing the temperature-, laser power–, time-, and surface ligand–dependence of both Stokes and anti-Stokes Raman intensities.

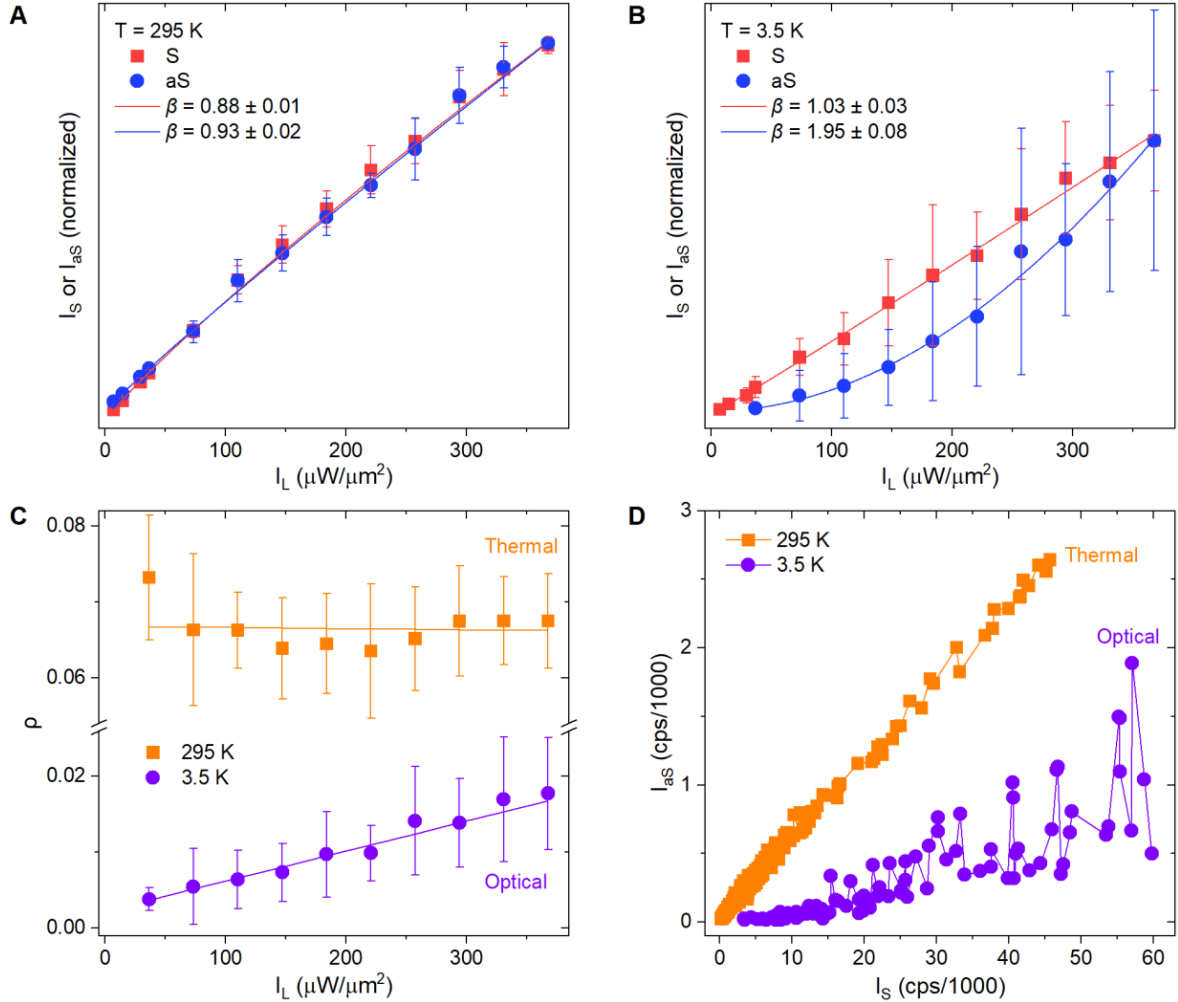

**Fig. 2.** Laser intensity and temperature dependence of Stokes and anti-Stokes Raman scattering. (**A, B**) Normalized Stokes (red squares) and anti-Stokes (blue circles) scattering peak intensities as a function of laser intensity at 295 K (**A**) and 3.5 K (**B**). The solid lines are fits to a power function, $y = a + bx^\beta$. (**C**) Anti-Stokes to Stokes intensity ratio ($\rho = I_{aS}/I_S$) at 295 K (orange squares) and 3.5 K (violet circles). (**D**) Anti-Stokes versus Stokes scattering intensities, showing thermal pumping at 295 K (orange squares) and optical pumping at 3.5 K (violet circles).

**Plasmon-Driven Vibrational Excitation at Room and Cryogenic Temperatures**

At room temperature (295 K), both the Stokes and anti-Stokes scattering intensities increase approximately linearly with laser intensity ($I_l$), exhibiting a slight sublinear deviation at higher intensities attributable to saturation effects, as evidenced by the exponent ($\beta$) values of the fitted power-law function (Fig. 2A & figs. S4-5). The slightly larger value of $\beta$ for the anti-Stokes scattering than for the Stokes (Fig. 2A) indicates the contribution of minor effect of vibrational optical pumping due to the particle plasmon excitation, which is in agreement with the calculated results (fig. S3c). However, the overall linear dependence underscores the dominance of the thermal population at 295 K. In contrast, at 3.5 K, the anti-Stokes intensity exhibits a quadratic dependence on laser power, whereas the Stokes intensity is linear (Fig. 2B and fig. S6), in very good agreement with predictions from plasmon-pumped vibrational population dynamics (*28*) and the quantum mechanical description of molecular optomechanics (*16, 29*) as observed trends are reproduced in the theoretical calculations (fig. S3D). The corresponding ratio ($\rho = I_{aS}/I_S$) plotted in Fig. 2C also exhibit distinct laser-power dependencies at 295 K and 3.5 K, confirming that different mechanisms dominate at the two temperatures. Over the given laser power range, the ratio remains approximately constant at 295 K, indicating that optical pumping of the vibrational population is negligible compared to thermal contributions. By contrast, at 3.5 K the ratio increases approximately linearly, consistent with optical pumping of the vibrational population (*28*). This behavior also agrees with the trends expected for weak plasmon–vibration optomechanical coupling, as confirmed by our calculated results (fig. S3C) based on the theoretical framework of molecular optomechanics (*29, 30*). The correlation between $I_{aS}$ and $I_S$ (Fig. 2D) further highlights the distinct behaviors of the anti-Stokes scattering intensity due to thermal pumping at 295 K and optical pumping at 3.5 K, which is also reproduced in the theoretical calculation (fig. S3F) based on the principles of molecular optomechanics.

In addition to the differences in functional dependence, linear at 295 K and quadratic at 3.5 K, the correlation plot (Fig. 2D) reveals markedly higher levels of signal fluctuation at 3.5 K (see also fig. S6 compared with figs. S4–S5). This fluctuation originates from the optically pumped vibrational population and reveals important surface dynamics. Because SERS is exquisitely sensitive to surface–molecule interactions, the observed fluctuations, which is most prominent at higher incident laser intensities, likely arise from laser- or plasmon-induced changes in adsorption geometry. At room temperature, the thermal energy (~25 meV) is sufficient to overcome potential barriers between local minima, enabling molecules to dynamically sample multiple configurations. By contrast, at 3.5 K, the thermal energy (~0.3 meV) is far too low to surmount such barriers, trapping molecules in specific adsorption states. Under these conditions, the adsorption configuration becomes highly sensitive to perturbation by laser excitation or local plasmonic fields, which can modulate both the SERS intensity and the efficiency of plasmon-pumping of vibrational population, giving rise to the observed fluctuations. Such fluctuations are absent for thionine adsorbed on citrate-stabilized gold nanoparticles (as it will be shown), indicating that the behavior arises from bromide-mediated molecular alignment and surface dynamics. These effects become even more evident in time-resolved SERS measurements at 295 K and 3.5 K, as discussed next.

**Time and Temperature Dependent Fluctuations and Excitation of Dark Modes**

Fig. 3A–B show intensity maps of spectra recorded at 1-second intervals, illustrating the signal evolution during the first 60 seconds (the behavior over extended illumination period at different laser powers are presented in figs. S7 and S8). At 295 K, both the Stokes and anti-Stokes signals remain stable, and no new vibrational bands are observed (Fig. 3A, C), confirming that the molecule does not undergo photochemical conversion. The temporal behavior of the anti-Stokes signal at 3.5 K (Figs. 3B, D, figs. S8A–B) is markedly different from that at 295 K (Figs. 3A, C, fig. S7), exhibiting pronounced signal fluctuations and new vibrational peaks over time. In the example shown in Fig. 3A-B, the anti-Stokes signal shows fluctuations up to ~1000% relative to its initial intensity, in stark contrast to the negligible fluctuations observed at 295 K as illustrated

in Fig. 3E. The different temporal behaviors depending on temperature are consistently reproduced across measurements taken at multiple locations (figs. S7 and S8A–B), with the low temperature anti-Stokes signal variations exceeding 1200% in some cases. The temperature-dependent fluctuations of the anti-Stokes signal, recorded at multiple locations and under varying laser powers, are summarized in Fig. 3F, where the anti-Stokes peak intensity is plotted against the corresponding Stokes peak intensity. At 295 K, the data exhibit a linear correlation (Fig. 3F, red circles). In contrast, at 3.5 K (blue lines), the anti-Stokes signal follows an overall quadratic dependence on the Stokes scattering intensity, superimposed with strong fluctuations, indicating significant signal instability under the cryogenic conditions. In some instances, the optically pumped anti-Stokes signal at 3.5 K exceed the thermally pumped signal at 295 K particularly at high laser intensity.

As mentioned above, blinking of the optically pumped anti-Stokes signal is observed when thionine is adsorbed on gold nanorods with CTAB surface ligand. CTAB-stabilized gold nanorods are known to retain strongly bound bromide ions (Br⁻) at the gold surface even after centrifugation and resuspension (*31*). Phenothiazinium dyes have also been shown to co-adsorb on gold nanorods together with Br⁻ (*32*), which can bridge surface–molecule electrostatic interactions (Fig. 3G). In addition, halides are known to induce local restructuring of surface gold atoms (*33-35*) that can lead to creation of metastable adsorption configurations as illustrated in Fig. 3G. Recent work also shows that halide adlayers undergo significant charge transfer with gold, generating outward-facing dipoles (*36*), that can guide molecular alignment in plasmonic nanocavity as shown in Fig. 3G. Au-Br⁻ bond exhibits a relatively strong Raman scattering peak at ~180 cm$^{-1}$ (*37*), and excitation of this vibrational mode may initiate displacement of a gold atom from its lattice site, creating atom protrusion. Such atomic-scale protrusions enhance both electromagnetic field localization and surface–molecule coupling, resulting in a sharp increase in the SERS signal (*38*). As the gold atom relaxes back to its equilibrium geometry ($i \leftrightarrow ii$, Fig. 3G), the enhanced signal vanishes. These dynamic surface processes provide plausible explanation for the observed SERS blinking. This bromide-induced molecular alignment and surface atom dynamics may explain the photocatalytic N-demethylation of methylene blue (*32*) and conversion of p-aminothiophenol to p-nitrothiophenol (*37*) on gold nanostructures in the presence of bromide.

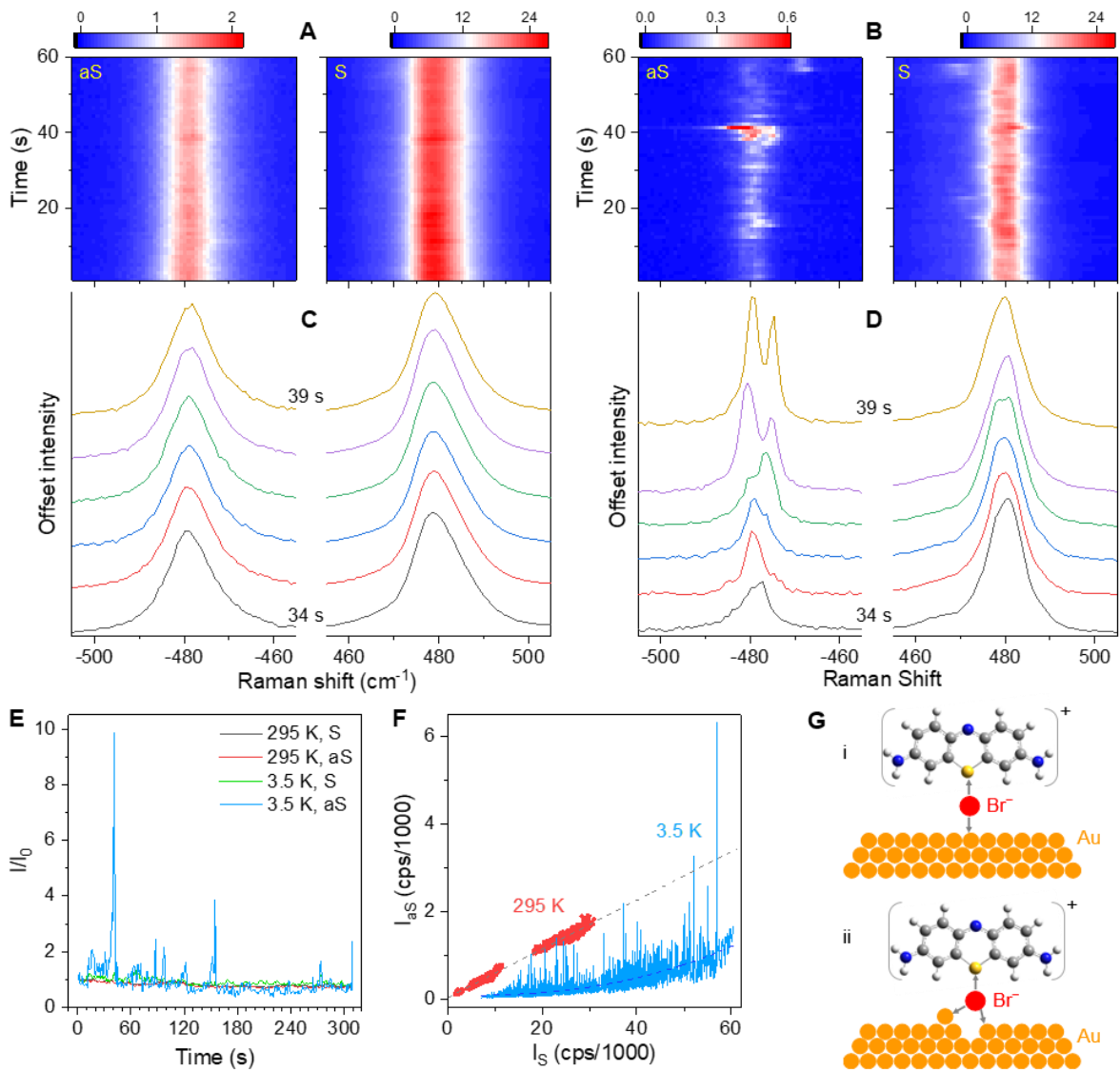

**Figure 3** Time dependence of the SERS signal at 295 K and 3.5 K. (**A**) Intensity map and (**B**) representative spectra at 295 K showing stable signal during 60 s of continuous illumination. (**C**) Intensity map and (**D**) representative spectra at 3.5 K showing pronounced fluctuations and multiple peaks over time. (**E**) Time dependence of peak intensity, normalized to the initial value, showing strong fluctuations only for $I_{aS}$ at 3.5 K. (**F**) $I_{aS}$ versus $I_S$, with a linear correlation at 295 K (red squares) and a quadratic correlation at 3.5 K (blue line) accompanied by strong fluctuations. (**G**) Schematic illustrating Br⁻-guided alignment of the thionine dipole and associated surface atom displacement.

Further insight into surface–molecule interactions and the presence of local energy minima emerges from the appearance of new spectral features at low temperature (Fig. 3D), most prominently in the anti-Stokes Raman bands. These multiple peaks are reproducibly observed across different spatial locations (Figs. 4B–F), indicating that they originate from distinct transient adsorption configurations. This behavior contrasts sharply with the broad, single peak observed at 295 K in both the Stokes and anti-Stokes branches (Figs. 3A and 4A), independent of spatial position or illumination time. The single room-temperature peak is consistent with theoretical calculations predicting a dominant Raman mode (Fig. 4G, pink line). At 295 K, the spectra reflect thermally averaged equilibrium configurations, with surface–molecule interactions playing a minimal role, as evidenced by the excellent agreement between experiment and calculations for the isolated molecule; a scaling factor of only 0.996 is sufficient to align the calculated and measured frequencies.

The multiple new peaks observed at 3.5 K, particularly well resolved in the anti-Stokes branch (Figs. 4B–F, blue traces), arise from the activation of otherwise "dark" or weakly Raman-active vibrational modes under specific surface–molecule adsorption geometries. As mentioned earlier, at low temperatures, molecules can become frozen in different adsorption configurations. Laser excitation and plasmon activation of the Au-Br⁻ bond can alter these configurations, thereby modifying the Raman polarizability tensor (*39*). For example, the calculated spectra (Fig. 4G) show vibrational modes at 473 cm$^{-1}$ (dashed orange line) and 463 cm$^{-1}$ (dashed green line), both with very weak Raman activity but moderately strong infrared absorbance. These modes are observed in the experimental SERS spectra and, in some cases, appear with even higher intensity than the 479 cm$^{-1}$ mode (Fig. 4E), which exhibits the strongest and most dominant Raman activity for the isolated molecule. Similarly, a vibrational mode at 440 cm$^{-1}$, predicted to be IR-active only and Raman-inactive, is observed in some of the SERS spectra (Figs. 4E–F) with a relatively strong scattering signal. Strong surface-molecule interaction can also shift vibrational frequencies, helping to explain the discrepancies between the observed peak positions (Figs. 4B–F) and the theoretical values (Fig. 4G). Overall, the anti-Stokes intensity for specific vibrational modes that appear and disappear over time provides evidence for vibrational pumping, driven by the synergistic effects of dynamical surface-molecule interaction and optomechanical plasmon-vibration coupling.

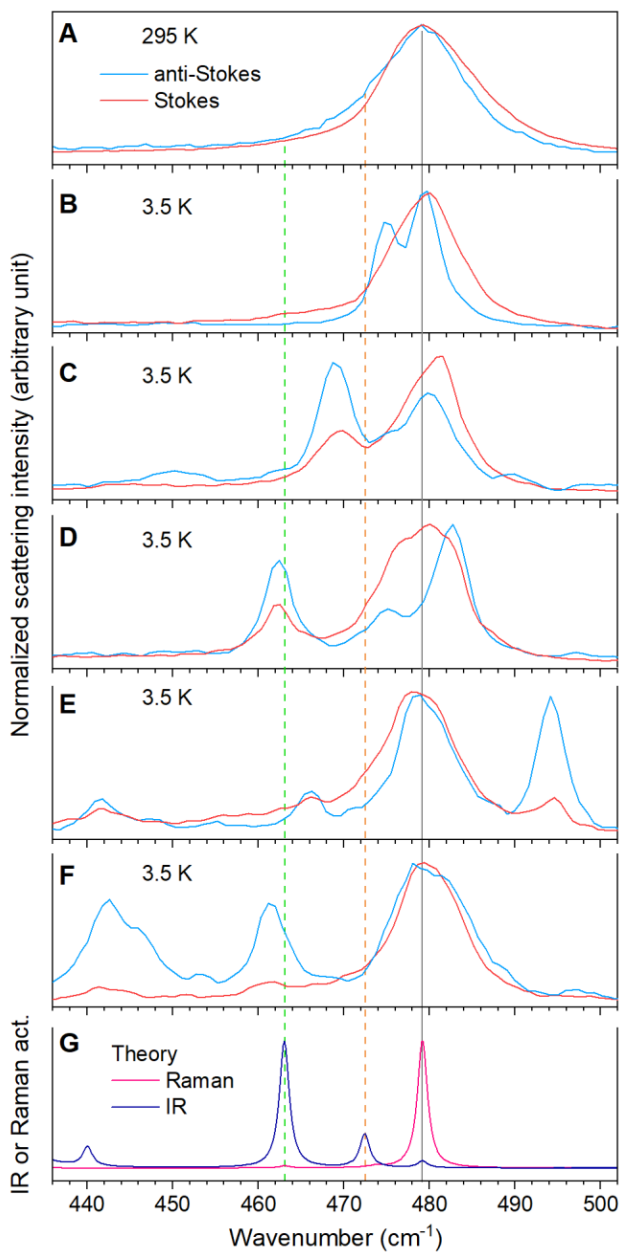

**Fig. 4 Representative spectra selected from the time series to illustrate the origin of multiple peaks at 3.5 K.** The Stokes and anti-Stokes spectra showing one broad band at 295 K (**A**), and multiple peaks at 3.5 K (**B-F**). (**G**) Calculated Raman spectrum (pink line) and IR spectrum (blue line).

**Vibrational Pumping in the Absence of Surface-Molecule Mediator**

Previous experimental studies have shown that phenothiazinium dyes, such as thionine and methylene blue, can replace carboxylic acid surface ligands and directly adsorb onto gold nanoparticles (*32*). In the absence of mediating co-adsorbates, theoretical calculations indicate that phenothiazinium dyes interact with the gold surface primarily through π-stacking–type dispersion forces (*22*), favoring a flat adsorption geometry (Fig. 1A, right). Interestingly, two types of SERS spectra are observed at 3.5 K experimentally (labeled type I and II in Fig. 5A) for thionine adsorbed on citrate-stabilized gold nanoparticles. Type I spectra exhibit weak Stokes and strong anti-Stokes scattering, whereas type II spectra show strong Stokes and weak anti-Stokes scattering. Intensity plots from multiple locations (Figs. 5B–C) confirm this dichotomy, which becomes more pronounced when $\rho$ is plotted against $I_L$ (Fig. 5D) and is most clearly resolved in the correlation plots of $I_{aS}$ and $\rho$ versus $I_S$ (Figs. 5E–F). Compared with type I, the intensities of type II spectra show a laser-power dependence that more closely resembles optical pumping of vibrational populations observed on gold nanorods. However, the power-law exponents (0.64±0.02 for $I_S$, 2.8±0.2 for $I_{aS}$ and 2.3±0.2 for $\rho$) deviate significantly from ideal behavior of optical pumping of vibrational population, suggesting additional contributions from charge-transfer processes. These contributions may arise from a more tilted adsorption geometry in the absence of a co-adsorbate that mediates molecular alignment. Nevertheless, the closer agreement of the type II trends with optical pumping predictions is consistent with the interaction Hamiltonian of molecular optomechanics, in which plasmon–vibration coupling is maximized when the molecular polarization and local electric-field vectors are aligned (*40*) as described in more detail in the Supplementary Materials.

The intensities of Type I spectra display markedly different laser power dependence from the optical pumping of vibrational population. The Stokes signal deviates drastically from linear trend, instead following first-order adsorption kinetics (see inset in Fig. 5B). $I_{aS}$ and $\rho$ exhibits sublinear behavior, with exponents of 0.7±0.1 and 0.5±0.1, respectively, in contrast to the quadratic dependence in type II spectra. Since the number of molecules in the focal volume remains approximately constant (assuming negligible diffusion), the observed signal saturation with increasing laser power can be attributed to changes in molecular orientation relative to the near-field vector. The much weaker Stokes signal for type I compared to type II (Fig. 5A) suggests that the molecular polarization has a smaller projection along the near-field, consistent with molecules lying approximately flat on the metal surface (Fig. 1A, left). Thus, the observed saturation may indicate a molecular reorientation toward a perpendicular alignment with the electric field ($\theta \rightarrow 90°$ in Fig. 1A). The correlation between weak Stokes and strong anti-Stokes signals in type I spectra clearly indicates that this orientation facilitates charge transfer, leading to efficient vibrational excitation, as reflected in the unusually large ratio $\rho$ (Fig. 5D). Type II spectra may originate from a minority population within the focal volume, given that polarization along the near field favors large Raman scattering signals.

For both spectral types on citrate-stabilized gold nanoparticles, the Stokes and anti-Stokes scattering intensities increase monotonically with laser power at 3.5 K (Figs. 5C and 5D), in stark contrast to the pronounced signal fluctuations observed for CTAB-stabilized gold nanorods (Fig. 2D and fig. S6). This difference highlights the critical role of bromide as a co-adsorbate. Consistently, negligible temporal fluctuations are observed in both the Stokes and anti-Stokes spectra on citrate-stabilized nanoparticles (Figs. 5G and 5H), whereas strong fluctuations emerge in the presence of bromide (Figs. 3E and 3F). Taken together, these results underscore the central role of bromide in directing molecular alignment and promoting surface atom dynamics. Such adsorption-mediated effects may be harnessed for photocatalytic applications.

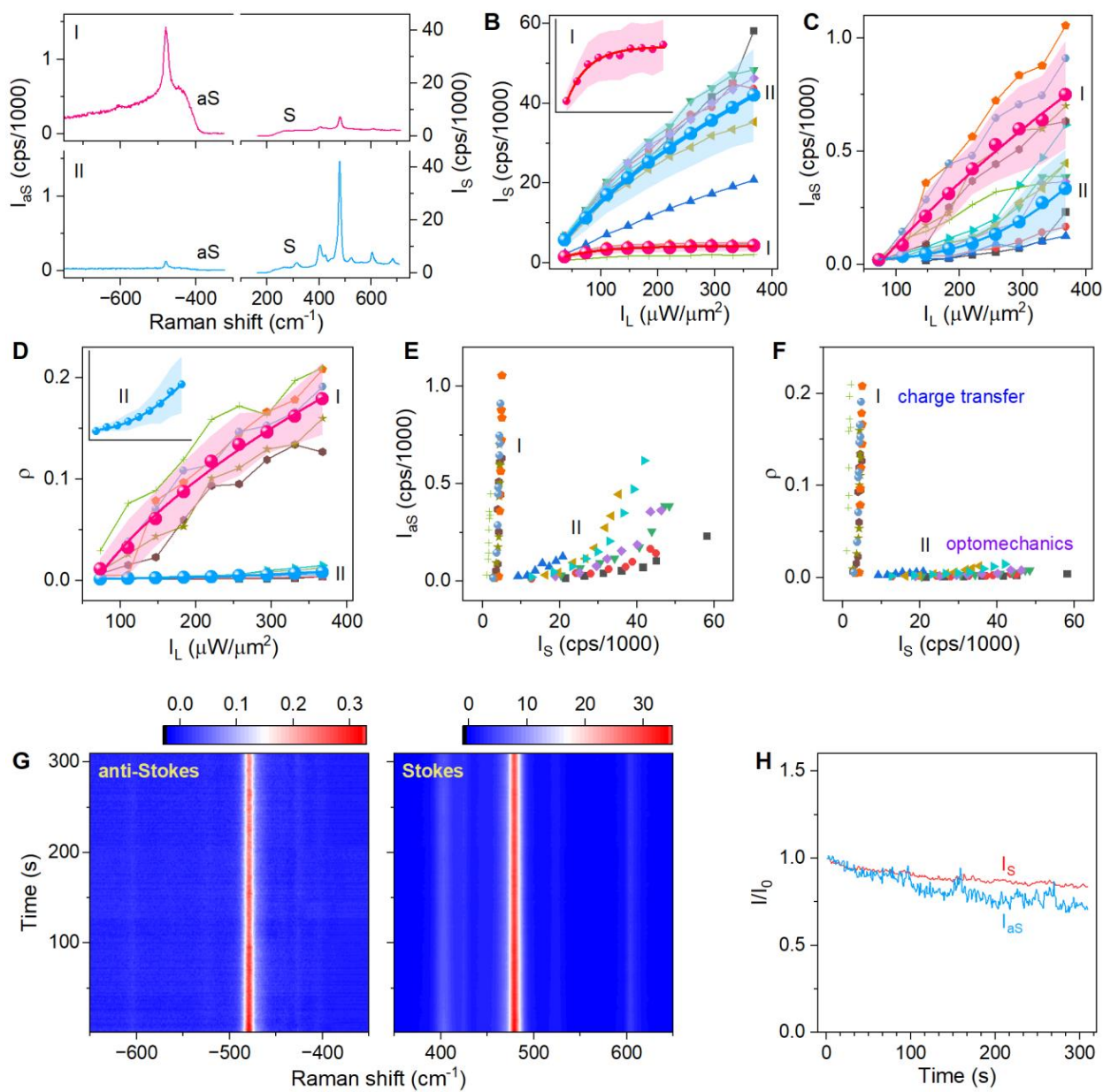

**Fig. 5 Laser power and time dependence of thionine SERS on citrate-stabilized gold nanoparticles at 3.5 K**. (**A**) Two types of SERS spectra: type I (top) with strong Stokes and weak anti-Stokes intensities, and type II (bottom) with weak Stokes and strong anti-Stokes intensities. (**B**) $I_S$ and (**C**) $I_{aS}$ versus laser intensity, showing reversed relative peak intensities for types I and II. (**D**) ρ as a function of laser intensity, distinguishing the two types. Large spheres denote averaged data. In (**B-C**) the large spheres represent the average data. (**E**, **F**) Correlation of $I_{aS}$ and ρ with $I_S$, further separating the two types. (**G**) Intensity map showing stability of Stokes and anti-Stokes bands. (**H**) Normalized peak intensities as a function of time.

In summary, using temperature-dependent SERS of appropriate probe molecule, we have disentangled coherent plasmon–vibration optomechanical coupling from incoherent hot-

electron–driven excitation of vibrational population. By comparing room-temperature and cryogenic measurements of thionine on gold nanostructures, we show that optical vibrational pumping dominates at 3.5 K, while thermal populations govern room-temperature behavior. Bromide co-adsorbates play a decisive role, guiding molecular alignment, inducing surface atom dynamics, and enabling transient adsorption configurations that activate otherwise Raman-dark vibrational modes. In the absence of bromide, distinct excitation regimes emerge that reflect competing optomechanical and charge-transfer pathways. These findings establish molecular optomechanics as a sensitive probe of surface–molecule interactions and reveal how anion-mediated surface dynamics can be harnessed to control energy flow at metal interfaces.

**References and Notes**

**Acknowledgments:** This research has been supported by the US National Science Foundation Grant 2217786. The cryostat optical system was acquired with support from the Defense University Research Instrumentation Program (DURIP) grant of an Air Force Office of Scientific Research, Award No. FA9550-22-1-0477. **AI assistance is used for proofreading to correct typos and grammatical errors using prompt "proofread this" typically one of a few sentences at a time.**

**Author contributions:**

Conceptualization: TGH

Methodology: BG, JD, WMT

Theoretical calculation and computation: BG

Funding acquisition: TGH

Project administration: TGH

Supervision: TGH

Writing – original draft: BG, TGH

Writing – review & editing: TGH

**Competing interests:** Authors declare that they have no competing interests.

**Data and materials availability:** All data, code, and materials used in the analysis must be available in some form to any researcher for purposes of reproducing or extending the analysis. Include a note explaining any restrictions on materials, such as materials transfer agreements (MTAs). Note accession numbers to any data relating to the paper and deposited in a public database; include a brief description of the data set or model with the number. If all data are in the paper and supplementary materials, include the sentence "All data are available in the main text or the supplementary materials."

**Supplementary Materials**

Materials and Methods

Supplementary Text

Figs. S1 to S9

Table S1

# Supplementary Materials for

# Distinguishing Hot-Electron and Optomechanical Pathways at Metal-Molecule Interfaces

Bing Gao[1], Jameel Damoah[1], Wassie M. Takele[1]†, and Terefe G. Habteyes[1]*

Corresponding author: habteyes@unm.edu

**The PDF file includes:**

    Materials and Methods
    Supplementary Text
    Figs. S1 to S9
    Table S1

**Materials and Methods**

Materials

Thionin acetate salt ($C_{12}H_9N_3S \cdot C_2H_4O_2$), also known as 3,7-diamino-5-phenothiazinium acetate and commonly referred to as thionine (the cationic dye $C_{12}H_9N_3S^+$), was purchased from Sigma-Aldrich. Gold nanospheres (AuNSs, ~60 nm diameter) stabilized with citrate (Cit) surface ligands and gold nanorods (AuNRs, ~40 nm diameter and ~92 nm length) stabilized with cetyltrimethylammonium bromide (CTAB) surface ligands were obtained from Nanopartz, Inc.

Sample Preparation and Characterizations

A 0.5 mL aliquot of gold nanoparticles (AuNSs or AuNRs) was diluted to a total volume of 1.5 mL with deionized (DI) water. The resulting aqueous solution was centrifuged at 6000 rpm for 5 min to remove excess surface ligands, after which the supernatant was discarded. The gold nanoparticles were then resuspended in 1.0 mL of an aqueous thionine solution ($2.0 \times 10^{-4}$ M) and incubated for at least 4 h to ensure complete adsorption of thionine molecules onto the nanoparticle surfaces. The dye–nanoparticle solution was subsequently centrifuged again at 6000 rpm for 5 min to remove unadsorbed dye molecules, and the supernatant was discarded. The resulting solid residue of gold nanoparticles with adsorbed thionine was resuspended in 50 µL of ultrapure water. Approximately 30 µL of this colloidal solution was drop-cast onto precleaned silicon substrates (~1 cm × 1 cm), which had been cleaned by successive sonication for 5 min each in acetone, isopropanol, and DI water. The drop-cast films were allowed to dry under ambient conditions for 2 h. The prepared nanostructure samples were then ready for use.

Transmission Electron Microscope (TEM) and Scanning Electron Microscope (SEM) Imaging

TEM imaging of the nanoparticles was performed using a JEOL 2010 microscope to obtain high-resolution images (fig. S1A). The aggregation behavior of thionine-treated gold nanoparticles on silicon substrates was examined by SEM, with representative images of AuNR samples shown in Fig. S1C.

UV-Vis Absorption and Dark-field Scattering Spectroscopy

Dark-field scattering images of thionine-treated nanoparticles on silicon substrates were obtained using a modified inverted microscope (Olympus GX51F5) equipped with a 100× dark-field objective (NA = 0.9), a 100 W halogen white-light source, and a CMOS camera (UC30, Olympus). Approximately 10% of the scattered light was directed to the CMOS camera to acquire dark-field images (fig. S1D), while the remaining ~90% was directed to a CCD spectrometer (IsoPlane spectrograph, Princeton Instruments) equipped with a back-illuminated deep-depletion CCD camera thermoelectrically cooled to −75 °C to acquire scattering spectra (Fig. S1E, blue line). The UV–Vis absorption spectrum of thionine in aqueous solution was recorded using a Shimadzu UV-2550 spectrophotometer (fig. S1E, black line) and compared with the absorption spectrum of thionine adsorbed on gold nanostructures (fig. S1E, red line), taken from Ref (26).

Low Temperature Spectroscopy

Temperature-dependent surface-enhanced Raman scattering (SERS) spectroscopy was performed using an optical closed-cycle cryostat (attoDRY800, attocube) equipped with a low-temperature-compatible apochromatic objective (NA = 0.81). The cryostat system allows temperature variation from 3.5 K to above room temperature. In this work, SERS spectra recorded at 3.5 K and 295 K were analyzed. The optical layout is shown schematically in fig. S1F. A beam splitter (BS) directs 10% of the excitation beam (λ = 632.8 nm) toward a second

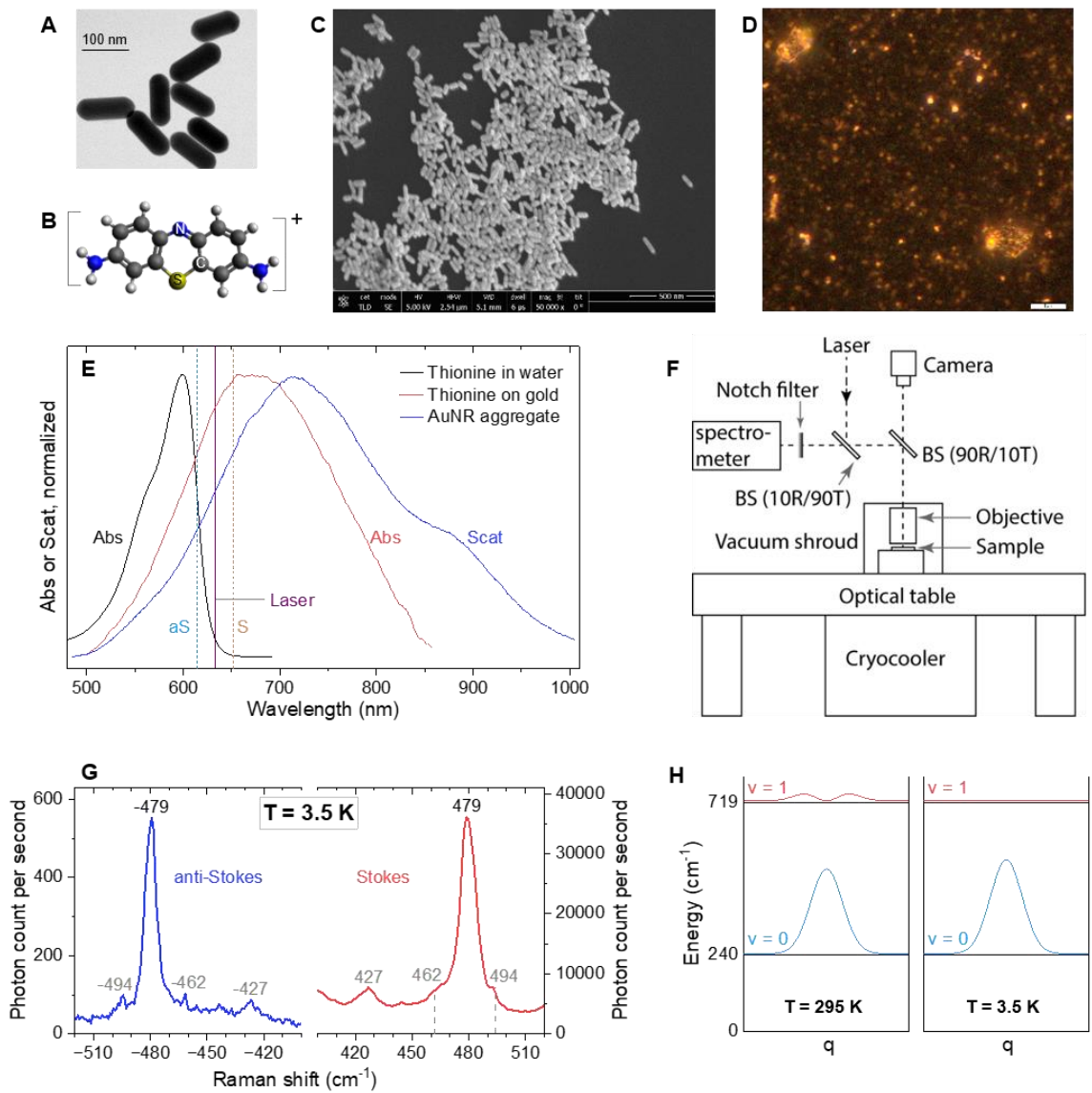

**Fig. S1. Sample and characterization.** (**A**) TEM image of AuNRs used as plasmonic systems. (**B**) Chemical structure of thionine. (**C**) SEM image of thionine-treated AuNRs on silicon substrate. (**D**) Dark-field image of thionine-treated AuNRs on silicon substrate. (**E**) Absorption spectrum of thionine in water solution (black line), on gold surface (red line; Ref. 1) and scattering spectrum of thionine-treated AuNR aggregates on silicon substrate (blue line). The purple vertical line indicates the excitation wavelength (632.8 nm). The dashed orange and blue lines indicate the Stokes and anti-Stokes shifts of the main vibrational band of thionine at 479 cm$^{-1}$. (**F**) Cryostat optical setup for temperature-dependent SERS measurement. (**G**) Anti-Stokes and Stokes SERS

spectra of thionine at 3.5 K showing a prominent peak at ±479 cm$^{-1}$. (**H**) Displacement vectors for the vibration mode at 479 cm$^{-1}$.

BS, which reflects 90% of the beam into the 0.81 NA objective that focuses the light onto the sample. The same objective is used to collect and collimate the scattered light from the sample. Approximately 90% of the collected signal is reflected by the BS closest to the sample, and 90% of this light is transmitted through the second BS and subsequently focused onto the entrance slit of the spectrometer after removal of the residual laser using a notch filter. The SERS spectra were recorded using an Andor spectrometer (Kymera 328i) equipped with a CCD camera (DU416A-LDC-DD), thermoelectrically cooled to −75 °C, with an acquisition time of 1.0 s and a grating of 1200 grooves mm$^{-1}$. Continuous variation of the incident laser power at the sample was achieved using a combination of two polarizers.

Figure 1G shows prominent anti-Stokes and Stokes SERS peaks at ±479 cm$^{-1}$ at 3.5 K. The atomic displacement vectors (Figure 1H, obtained using density functional theory calculation at B3LYP/aug-cc-pVDZ level of theory) indicate that the 479 cm$^{-1}$ mode corresponds to a skeletal deformation, in which distortion of the central ring containing the S and N atoms induces in-plane atomic displacements.

**Supplementary Text**

Resonance Raman effect

    Comparison of the calculated and experimental spectra indicates that electronic resonance Raman effects make a significant contribution to the thionine SERS spectra. Vibrational frequency calculations were performed using density functional theory (DFT) at the B3LYP/aug-cc-pVDZ level of theory with the Gaussian software package (Gaussian 16), utilizing the supercomputing facilities of the Center for Advanced Research Computing, University of New Mexico. Geometry optimization and frequency calculations were carried out for the cationic thionine (fig. S1B) without inclusion of the counterion. Resonance Raman spectra were simulated based on time-dependent density functional theory (TD-DFT) calculations using the same functional and basis set. A solvent model (SCRF = water) was employed for comparison with the experimental spectra, as it effectively captures solvation effects for charged molecular species. To obtain the resonance Raman spectrum, the ground and excited states of the thionine molecule were reoptimized using the Franck–Condon (FC) and Franck–Condon/Herzberg–Teller (FCHT) methods. The resonance Raman intensities were then derived from the calculated Raman activities according to the following relationship (41):

$$I_i = \frac{f(\tilde{v}_0 - \tilde{v}_i)^4 S_i}{\tilde{v}_i \left[1 - e^{\frac{-hc\tilde{v}_i}{kT}}\right]} \quad (1)$$

where:
    $f$ is a suitably chosen common normalization factor, $10^{-12}$, for the intensity of all peaks.
    $\tilde{v}_0$ is the laser exciting wavenumber in cm$^{-1}$
    $\tilde{v}_i$ is the vibrational wavenumber of the i$^{th}$ normal mode in cm$^{-1}$
    $T$ is temperature in Kelvin
    $S_i$ is the Raman scattering activity of the i$^{th}$ normal mode $v_i$
    $h$, $c$, and $k$ are Planck's constant, speed of light, Boltzmann constant, respectively

The calculated spectra are shown in fig. S2A (without resonance Raman effects) and fig. S2B (including resonance Raman effects) after applying Lorentzian line-shape broadening with a full width at half-maximum of 15 cm⁻¹. Notably, the experimentally measured SERS spectrum of thionine (fig. 2C) shows much closer agreement with the calculated spectrum that includes resonance Raman effects (fig. S2B).

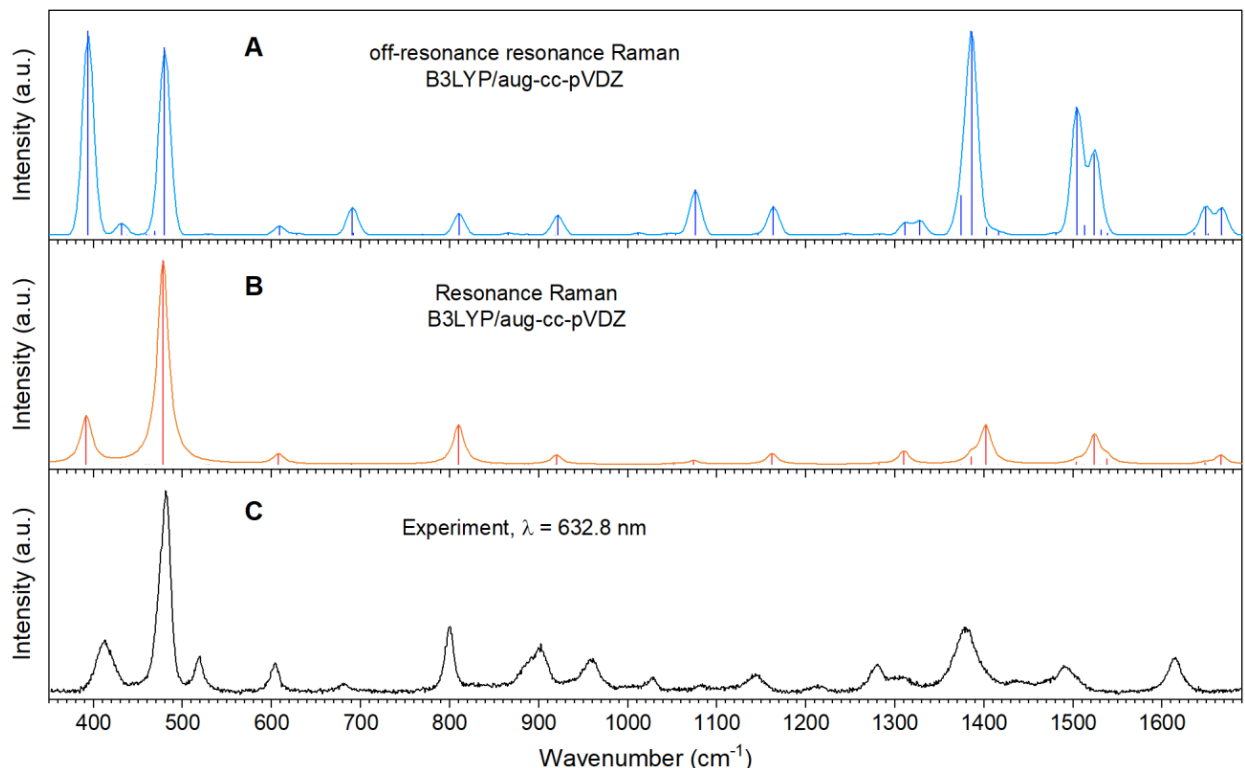

**Fig. S2. Evidence for resonance Raman effects in the thionine spectrum**. (**A**, **B**) Raman spectra calculated using the B3LYP functional with the aug-cc-pVDZ basis set (**A**) without and (**B**) with inclusion of resonance Raman effects. (**C**) The experimentally measured SERS spectrum of thionine shows much closer agreement with the calculated spectrum that includes resonance Raman effects (**B**).

Relevance of Thermal Pumping

According to the Boltzmann distribution, the anti-Stokes ($I_{aS}$) to Stokes ($I_S$) scattering intensity ratio ($\rho$), assuming comparable Raman polarizabilities, is given by

$$\rho = \frac{I_{aS}}{I_S} = \frac{(\tilde{v}_0+\tilde{v}_i)^4}{(\tilde{v}_0-\tilde{v}_i)^4} \exp\left(-\frac{hc\tilde{v}_i}{kT}\right) \qquad (2)$$

For $\tilde{v}_i = 479 \, cm^{-1}$, the ratio is plotted in Fig. 1C, and becomes negligible for $T < 50$ K, indicating the absence of thermal excitation, as the spacing between the vibrational energy levels becomes much larger than the thermal energy ($hc\tilde{v}_i \gg kT$).

The anti-Stokes scattering intensity observed at 3.5 K cannot be explained by an increase in local temperature due to plasmonic heating. If an increase in the vibrational population were driven by local plasmon heating, the anti-Stokes–to–Stokes intensity ratio would similarly increase with incident laser power at different initial temperatures. The ratio ($\rho$) is described by the Boltzmann distribution, as discussed above. For $\tilde{v}_i = 479 \, cm^{-1}$, the corresponding vibrational temperature ($\theta_v$) is given by $\theta_v = hc\tilde{v}/k = 689 \, K$, which simplifies the above relation to

$$\rho = 1.2745\exp\left(-\frac{\theta_v}{T}\right) \qquad (3)$$

The local temperature increase ($\Delta T$) due to plasmonic heating within the SERS hotspot can be accounted for by modifying the above equation as

$$\rho = 1.2745 \exp\left(-\frac{\theta_v}{T_i + \Delta T}\right) \qquad (4)$$

where $\Delta T$ is proportional to the absorption cross-section ($\sigma_{abs}$) and incident laser intensity ($I_l$) as $\Delta T \propto \sigma_{abs} I_l$. We have observe values of $\rho$ as high as 0.11 at an initial temperature $T_i = 3.5$ K. If this increase were attributed solely to local plasmonic heating, the corresponding temperature rise would be $\Delta T = 277$ K, yielding an effective temperature of $T = T_i + \Delta T = 280$ K. Such a temperature increase is unrealistic under our experimental conditions, where the incident laser power is low, with a maximum laser intensity below 370 µW/µm². Using the anti-Stokes–to–Stokes intensity ratio in SERS, Park et al.(42) reported a maximum temperature increase of approximately 60 K in gold nanoparticle aggregates at incident laser intensities exceeding 2000 µW/µm², which is more than fivefold higher than the intensity used in our experiments. In addition, it is highly unlikely that a local steady-state temperature could reach 280 K while the global sample temperature remains at 3.5 K.

Vibrational pumping

The observation of the anti-Stokes Raman scattering signal at low temperature (e.g. 3.5 K) can be explained in terms of vibrational optical pumping in SERS hotspots (28). According to the vibrational pumping theory, the laser intensity ($I_l$) dependence of Stokes scattering intensity, the anti-Stokes scattering intensity and their ratio are given by (28)

$$I_S = (1+n) \cong N\sigma_S I_l \qquad (5)$$

$$I_{aS} = nN\sigma_{aS} I_l = \left[\frac{\tau\sigma_S}{hc\tilde{v}_l} I_l + \exp\left(-\frac{hc\tilde{v}_v}{kT}\right)\right] N\sigma_{aS} I_l \qquad (6)$$

$$\rho = \frac{\sigma_{aS}}{\sigma_S}\left[\frac{\tau\sigma_S}{hc\tilde{v}_l} I_l + \exp\left(-\frac{hc\tilde{v}_v}{kT}\right)\right] \qquad (7)$$

where $\sigma_S$ and $\sigma_{aS}$ are Stokes and anti-Stokes Raman scattering cross-sections, respectively, $N$ is the number of molecules in the SERS hotspot, $\tau$ is the vibrational lifetime and $n$ is the Bose-Einstein occupation (vibrational Bose factor), $n \ll 1$. The relationships show that $I_S$ and $\rho$ depends linearly on $I_l$, whereas $I_{aS}$ exhibits a quadratic dependence. The trends observed in our experiments (Fig. 2) are consistent with these theoretical predictions.

Molecular optomechanics

Molecular optomechanics provides a general quantum-mechanical framework for describing plasmon–vibration interactions. In this approach, surface-enhanced Raman scattering (SERS) is modeled by quantizing both the molecular vibrational modes and the electromagnetic field of the plasmonic cavity as bosonic excitations (29, 30). Building on these theoretical frameworks, we show that our experimental observations are quantitatively reproduced within this quantum-mechanical description, demonstrating that the conventional theory of vibrational pumping corresponds to the weak-coupling limit of molecular optomechanics.

In molecular optomechanics theory, the total Hamiltonian of the laser-driven plasmon-molecule system given as

$$\hat{H} = \hat{H}_p + \hat{H}_v + \hat{H}_{int} \qquad (8)$$

where, the $\hat{H}_p$ and $\hat{H}_v$ are the Hamiltonian of the plasmon and molecular vibration, respectively, and both are modeled as harmonic oscillator. They are defined in terms of the bosonic annihilation ($\hat{a}$ or $\hat{b}$) and creation ($\hat{a}^\dagger$ or $\hat{b}^\dagger$) operators as $\hat{H}_p = \hbar\omega_p \hat{a}^\dagger \hat{a}$ and $\hat{H}_v = \hbar\omega_v \hat{b}^\dagger \hat{b}$, where $\omega_p$ and $\omega_v$ are the resonance angular frequencies of the plasmon and vibration oscillators, respectively. $\hbar$ is the reduced Planck's constant given as $\hbar = \frac{h}{2\pi}$.

The interaction Hamiltonian is given as

$$\hat{H}_{int} = -\hbar g_0 \hat{a}^\dagger \hat{a}(\hat{b} + \hat{b}^\dagger) \qquad (9)$$

where $g_0$ is the single plasmon optomechanical coupling rate that depends on the zero-point normal mode coordinate ($Q_v^0$), Raman activity ($R_v$) of the vibrational mode, the effective mode volume ($V_m$) of the plasmon field, and permittivity of the medium ($\varepsilon$) as shown by the relationship.

$$g_0 = \frac{Q_v^0 \omega_c \sqrt{R_v}}{2\varepsilon_0 \varepsilon V_m} (\mathbf{u_E} \cdot \mathbf{u_p})^2 \tag{10}$$

where the Raman activity is related to change in the molecular polarizability ($\alpha$) with respect to the normal coordinate, $R_v = (\partial \alpha / \partial Q_v)^2$. $\mathbf{u_E}$ and $\mathbf{u_p}$ are unit vectors of plasmon field and molecular polarization, respectively. That is, equation (10) is reduced to

$g_0 = \frac{Q_v^0 \omega_c \sqrt{R_v}}{2\varepsilon_0 \varepsilon V_m}$   for the molecular polarization parallel to the field

$g_0 = 0$   for the molecular polarization perpendicular to the field

The total Hamiltonian including the excitation laser can then be written as (40)

$$\hat{H} = \hbar\omega_p \hat{a}^\dagger \hat{a} + \hbar\omega_v \hat{b}^\dagger \hat{b} - \hbar g_0 \hat{a}^\dagger \hat{a}(\hat{b} + \hat{b}^\dagger) + i\hbar\Omega(\hat{a}^\dagger e^{-i\omega_l t} - \hat{a} e^{i\omega_l t}) \tag{13}$$

where $\Omega$ is pumping rate, which is related to the laser intensity as described in Ref. (40). It suffices to note that $\Omega^2 \propto I_l$, where $I_l$ is the laser intensity. Using this Hamiltonian, the master equation governing the dynamics of the system's density matrix [Eq. (2) in Ref. (30)] is solved numerically, following the approximation methods described in Ref. (30). The relevant results, obtained using the parameters listed in Table S1, are shown in Fig. S3 and can be directly compared with the experimental trends observed in Fig. 2.

Table S1. Summary of parameters used in the molecular optomechanics model

| Items | Values | | Description |
|---|---|---|---|
| | Experiments | Optomechanics model | |
| Vibrational mode | 479 cm$^{-1}$ | 0.05938 eV | $\omega_m$, the major peak of thionine |
| Photon decay rate | 15 cm$^{-1}$ | 0.00186 eV | $\gamma_m$, calculated from the experimental peak of thionine at 479 cm$^{-1}$ |
| Laser wavelength | 632.8 nm | 1.95872 eV | $\omega_L$ |
| Resonant frequency of the nanocavity | 730 nm | 1.69856 eV | $\omega_c$, calculated from dark-field scattering from aggregated AuNRs |
| Nanocavity damping rate | 95 nm | 0.25 eV | $\kappa$, calculated from dark-field scattering from single AuNR |
| Optomechanical coupling rate | / | 0.002 eV | $g$ |
| Laser power | 0.1-1.0 mW | 0.001-1.0 eV$^2$ | $\Omega^2$, there is no strict correspondence to the real laser power |
| Temperature | 3.5 K, 295 K | 3.5 K, 295 K | T |
| Detection system coefficient | / | 1 | $\alpha_{det}$, it should differ from the actual value but should not affect the anti-Stokes/Stokes ratio |

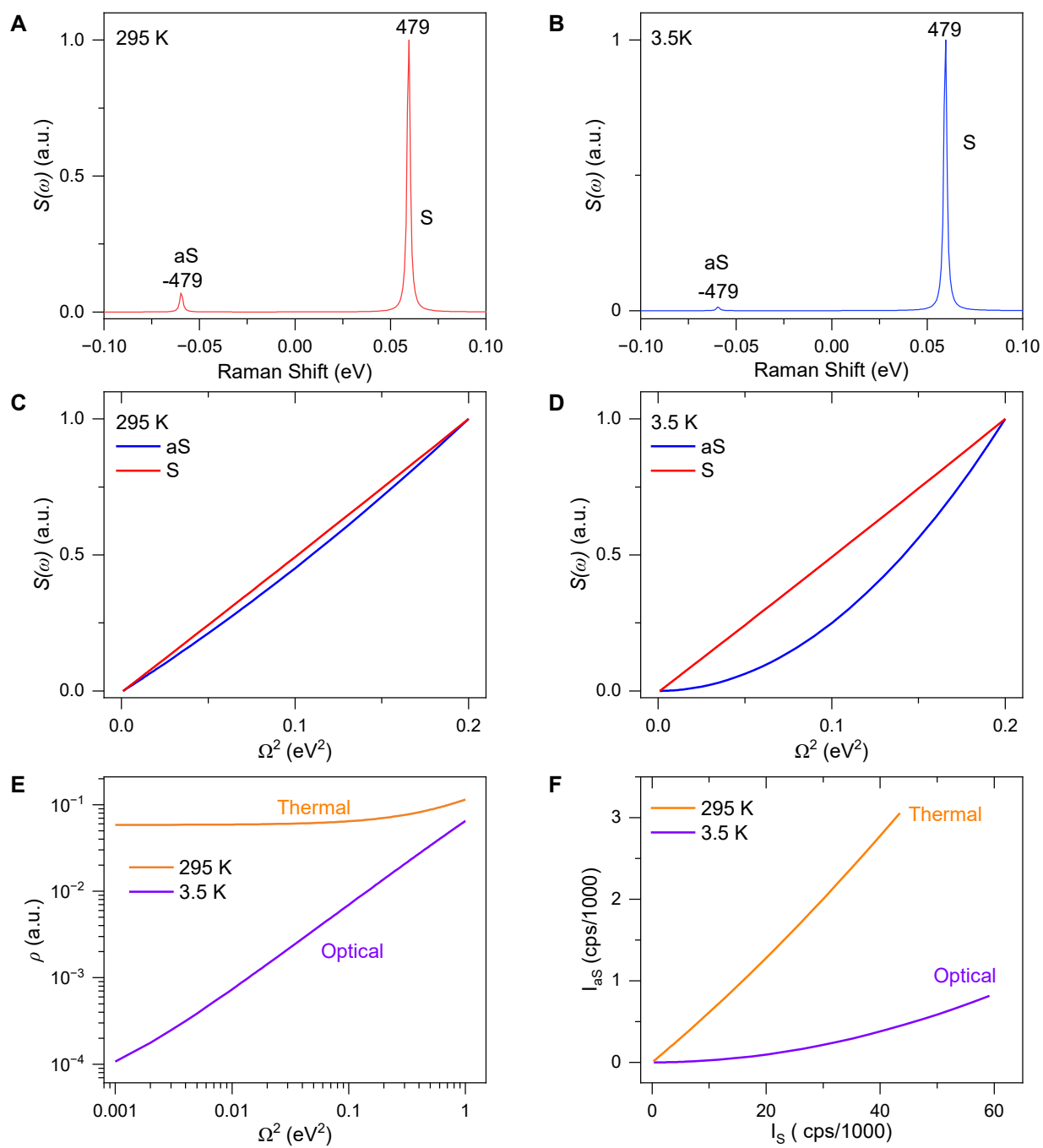

**Fig. S3. Molecular optomechanics**. (**A**, **B**) Calculated Stokes and anti-Stokes Raman scattering spectra at 295 K (**A**) and 3.5 K (**B**). (**C**, **D**) Laser pumping rate dependence of Stokes (red line) and anti-Stokes (blue line) scattering intensity at 295 K (**C**) and 3.5 K (**D**). (**E**) Anti-Stokes to Stokes intensity ratio ($\rho$) as a function of laser pumping rate. (**F**) Correlation between anti-Stokes and Stokes Raman scattering intensity.

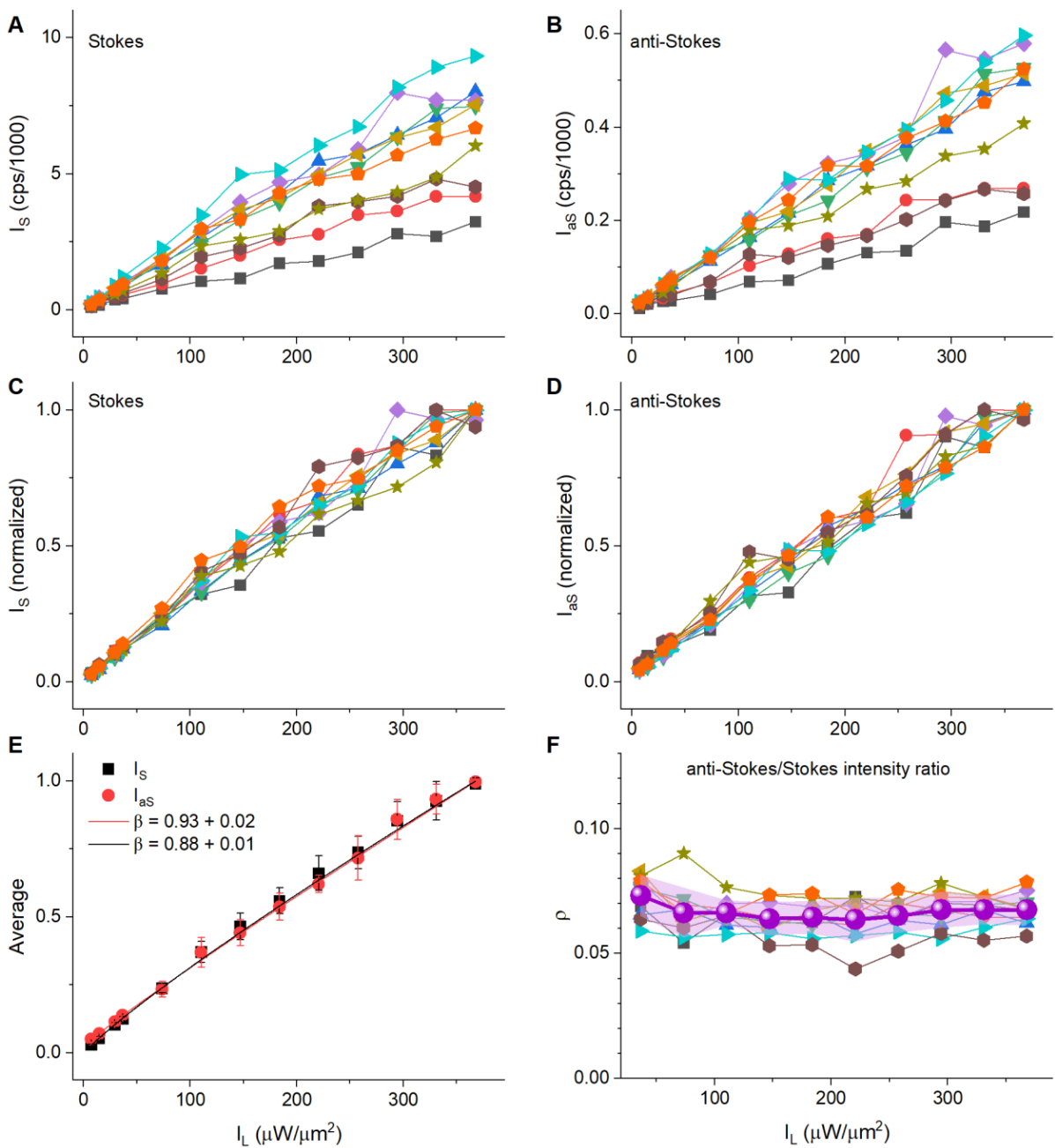

**Fig. S4. T = 295 K in vacuum. Laser intensity and temperature dependence of vibrational population at individual spots.** (A, B) Peak photon counts per second (cps) as a function laser intensity ($I_L$) for Stokes (A) and anti-Stokes scattering at 295 K. (C) Anti-Stokes to Stokes ratio ($\rho = I_{aS}/I_S$) at 295 K. (D) $I_{aS}$ as a function of $I_S$ at 295 K. (E-H) same as (A-D) but at 3.5 K.

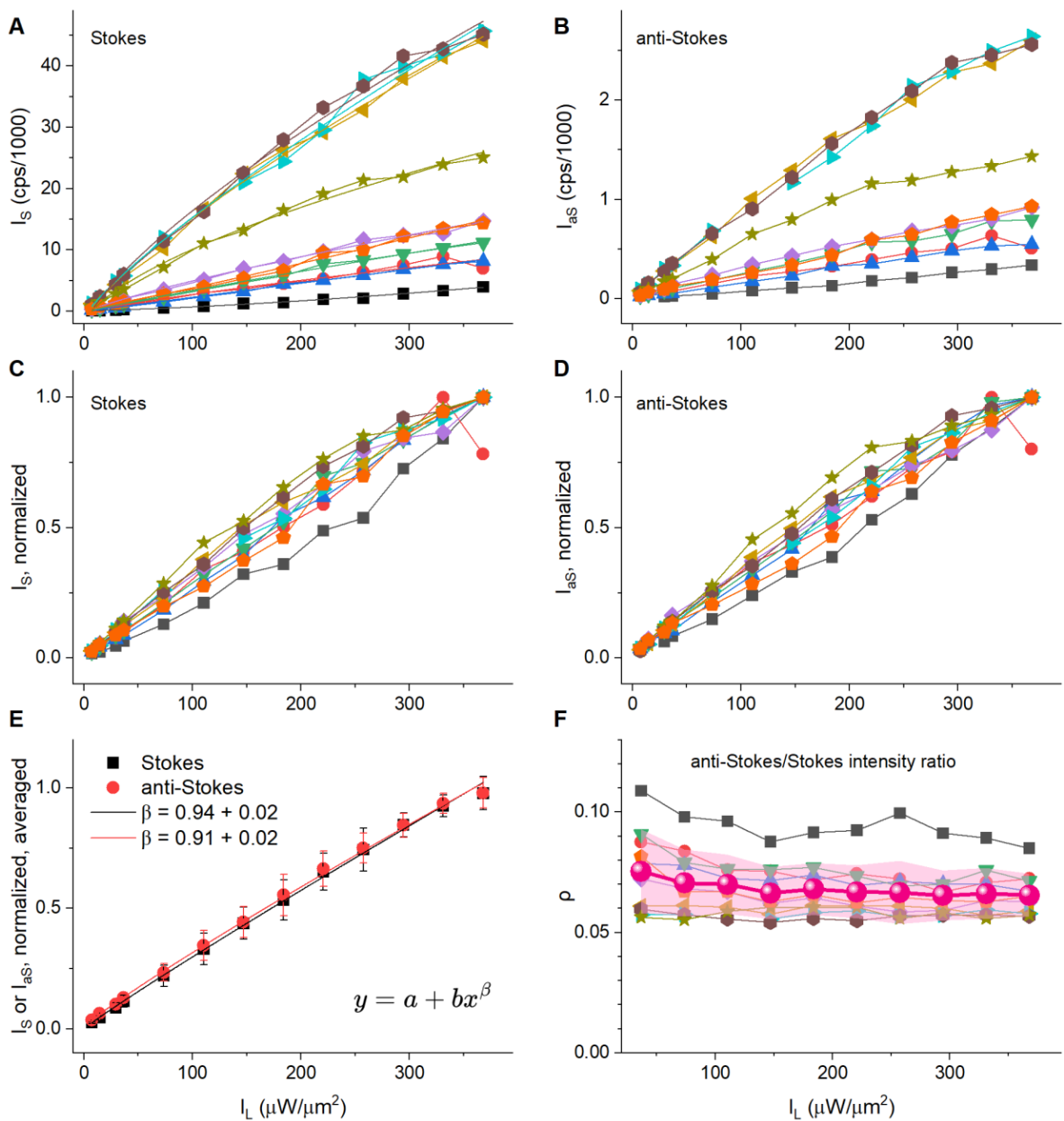

**Fig. S5. T = 295 K in air. Laser intensity and temperature dependence of vibrational population at individual spots.** (A, B) Peak photon counts per second (cps) as a function laser intensity (IL) for Stokes (A) and anti-Stokes scattering at 295 K. (C) Anti-Stokes to Stokes ratio (ρ=I_aS/I_S) at 295 K. (D) I_aS as a function of I_S at 295 K. (E-H) same as (A-D) but at 3.5 K.

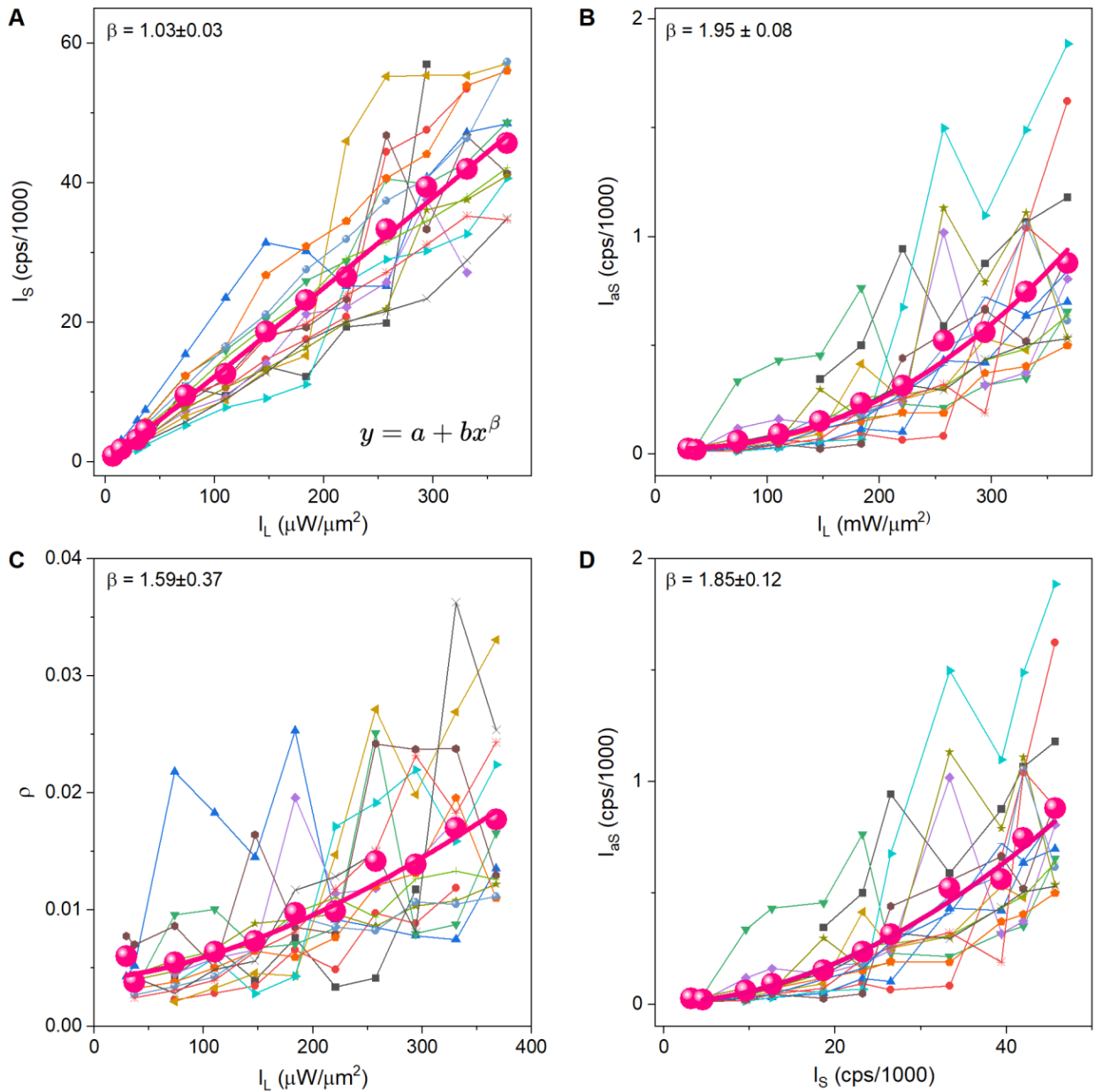

**Figure S6. T = 3.5 K. Laser intensity and temperature dependence of vibrational population at individual spots.** (A, B) Peak photon counts per second (cps) as a function laser intensity ($I_L$) for Stokes (A) and anti-Stokes scattering at 295 K. (C) Anti-Stokes to Stokes ratio ($\rho = I_{aS}/I_S$) at 295 K. (D) $I_{aS}$ as a function of $I_S$ at 295 K. (E-H) same as (A-D) but at 3.5 K.

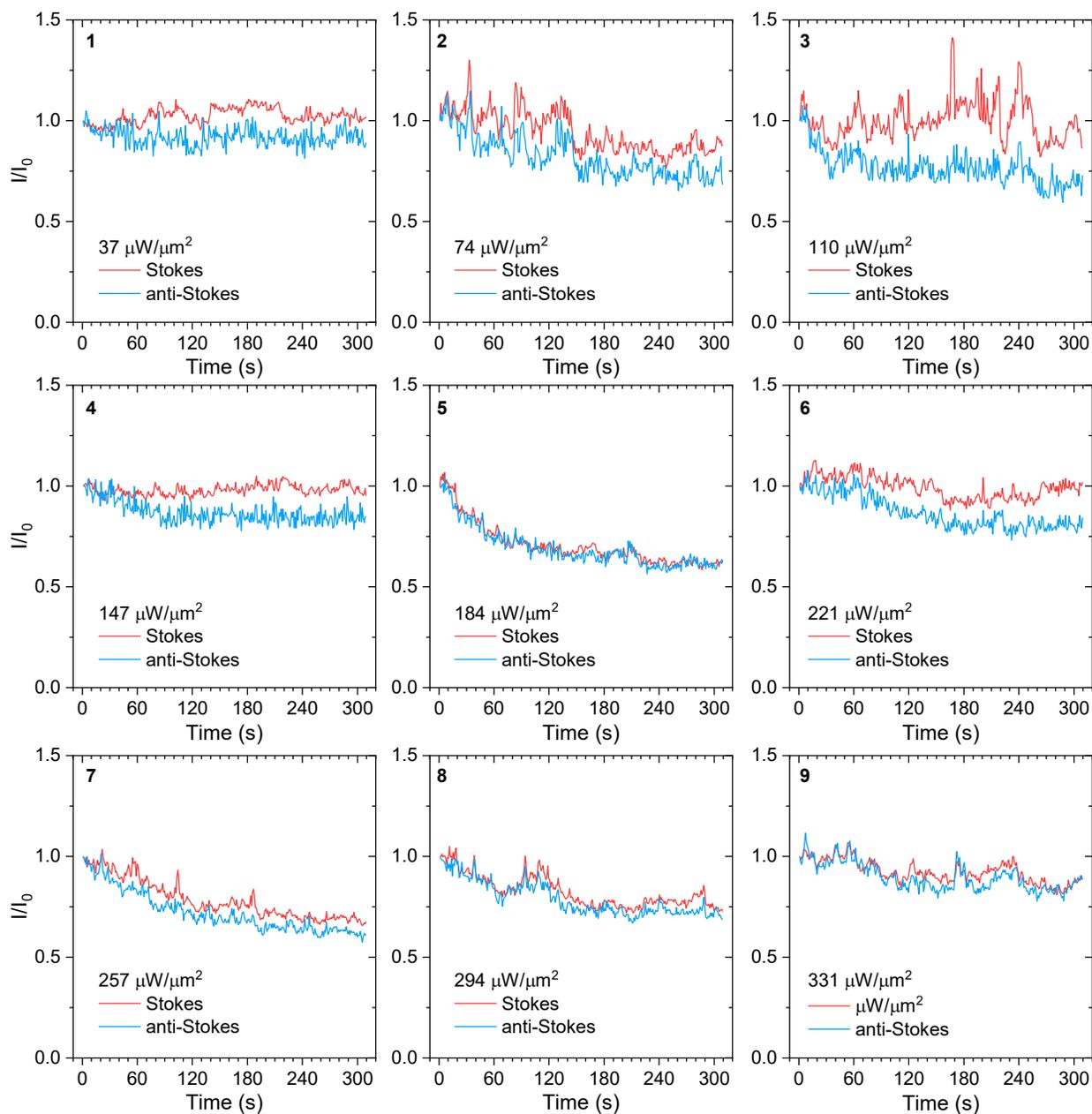

**Figure S7. Time dependence of the SERS signal at 295 K.** (1) Time dependence of peak intensity, normalized to the initial value at 295 K during 300 s of continuous illumination of $I_L$ = 37 µW/µm². (2-9) same as (1) but at different laser intensity from 74 to 331 µW/µm².

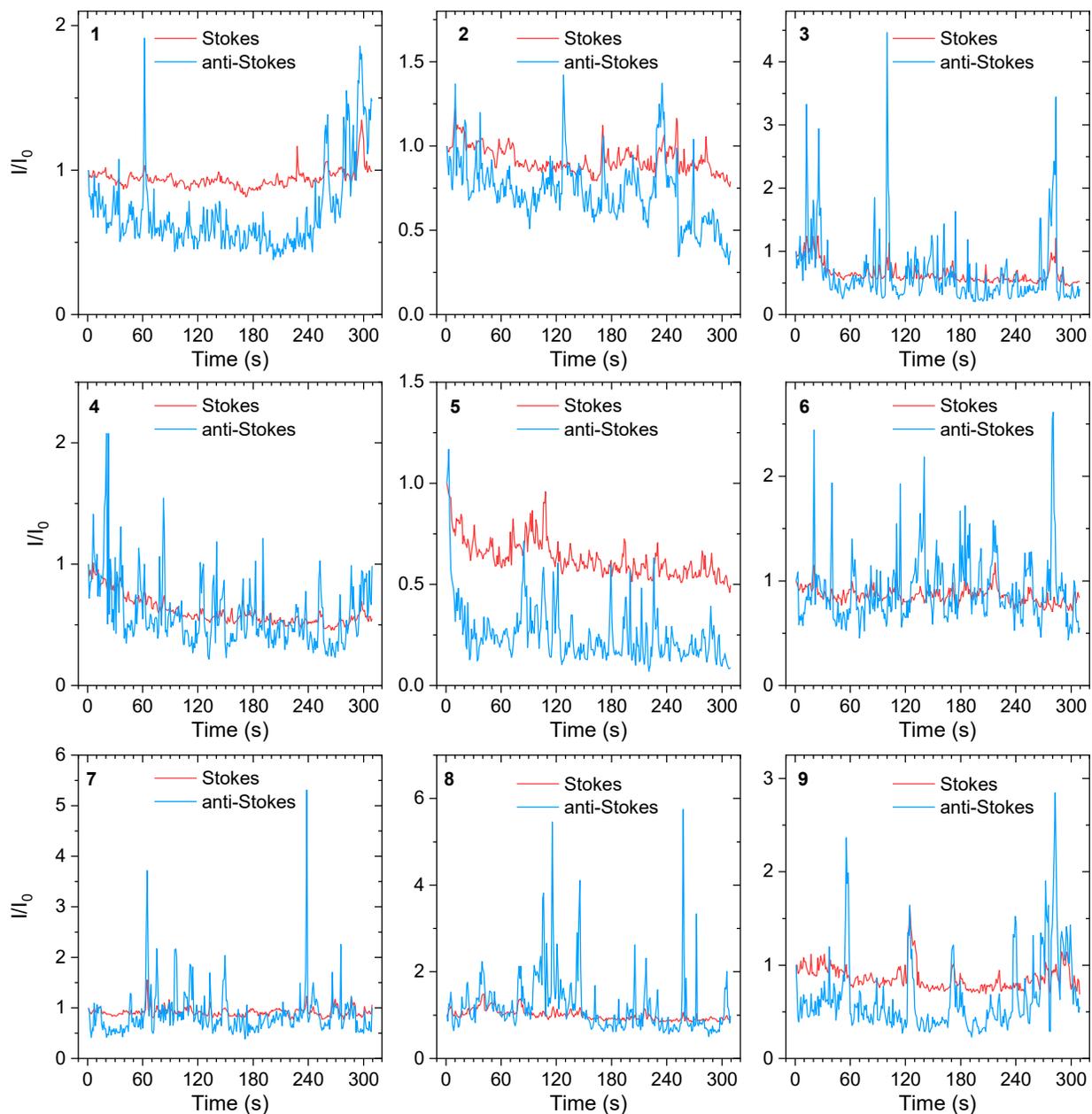

**Fig. S8A. Time dependence of the SERS signal at 3.5 K.** (1-9) Time dependence of peak intensity, normalized to the initial value at 3.5 K during 300 s of continuous illumination of $I_L = 147\ \mu W/\mu m^2$.

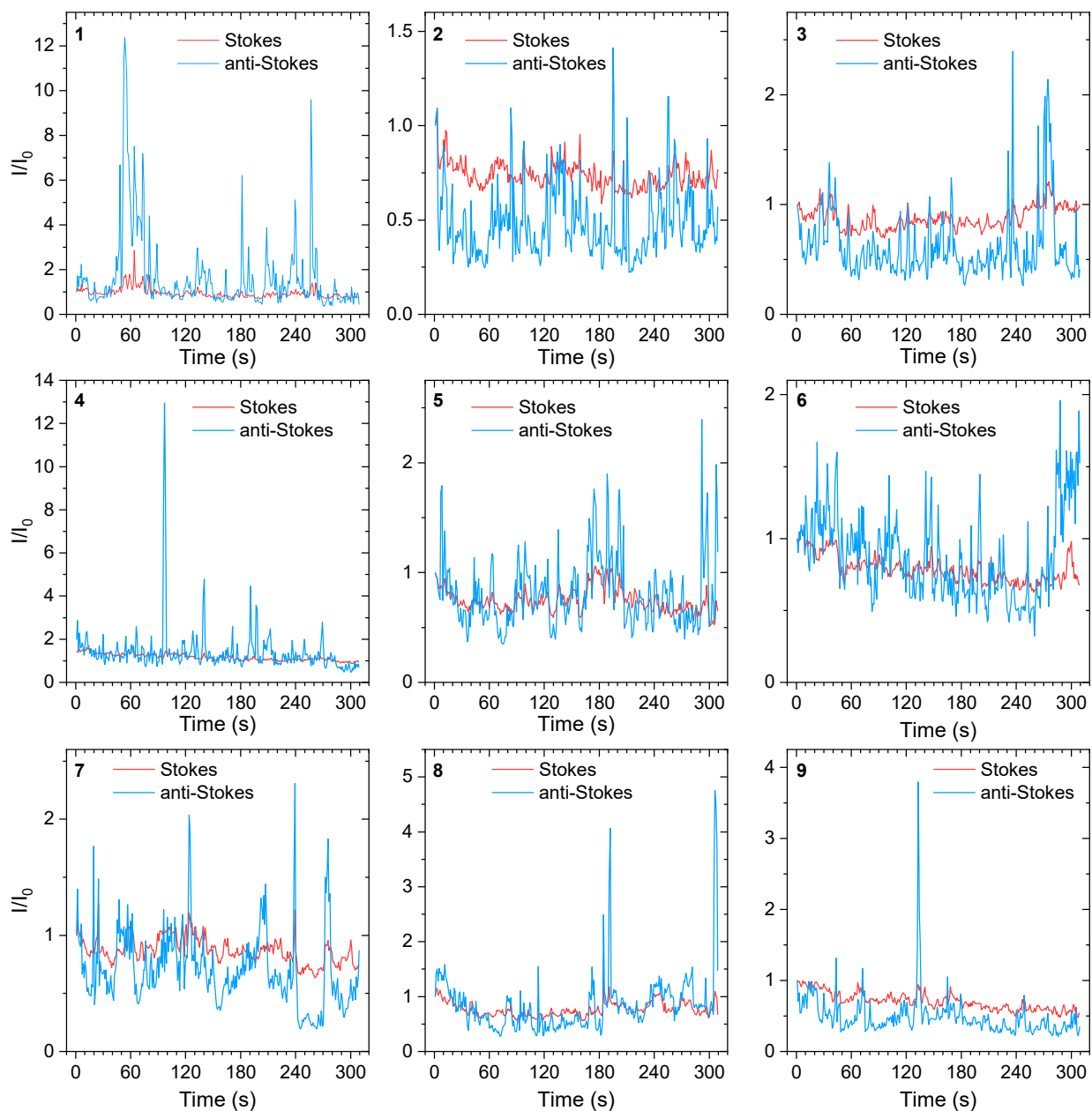

**Fig. S8B. Time dependence of the SERS signal at 3.5 K.** (1-9) Time dependence of peak intensity, normalized to the initial value at 3.5 K during 300 s of continuous illumination of $I_L = 294$ μW/μm$^2$.

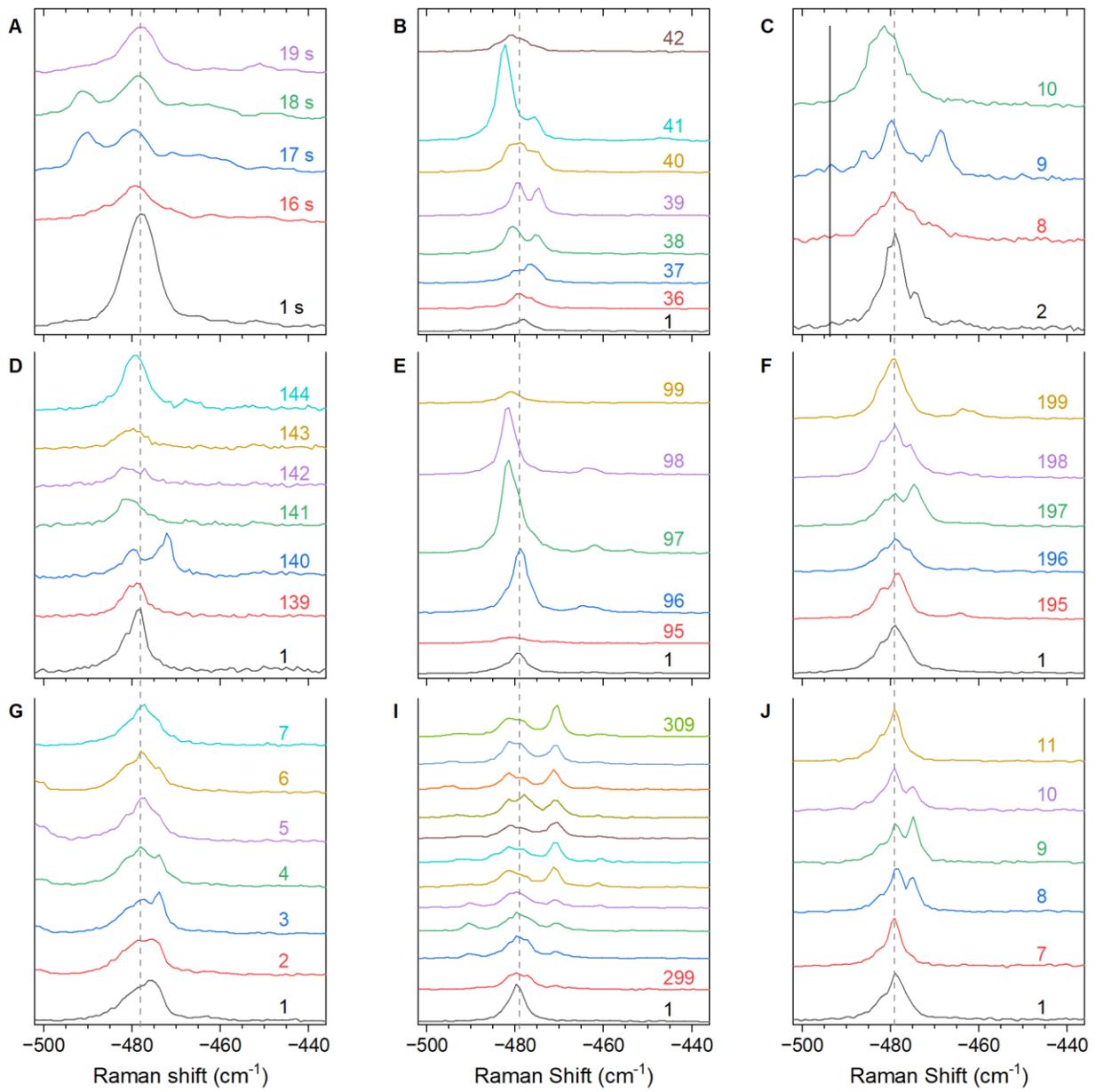

Fig. S9. Representative spectra selected from the time series in the anti-Stokes branch at 3.5 K. (a-j) Multiple peaks observed depending on location and illumination time, as labeled, at 3.5 K.